\documentclass{aa}  

\usepackage{graphicx}
\usepackage{txfonts}
\usepackage{lipsum}
\usepackage{lscape}             
\usepackage{placeins}           
\usepackage[table,xcdraw]{xcolor}
                                
\usepackage{hyperref}
\hypersetup{
    colorlinks=true,
    linkcolor=blue,
    filecolor=magenta,      
    urlcolor=cyan,
    linkcolor=blue,  
    citecolor=blue,        
}
\usepackage{orcidlink}

\usepackage{graphicx}	
\usepackage{amsmath}	
\usepackage{soul} 
\usepackage{booktabs}

\usepackage{titlesec}
\titleformat{\paragraph}[block]{\normalfont\normalsize\bfseries}{\theparagraph}{0em}{}
\titlespacing{\paragraph}{0pt}{2ex}{1ex}

\makeatletter
\newenvironment{dataavailability}{%
  \par\small\vspace{0ex}\noindent
  {\itshape Data Availability.}\enspace\ignorespaces
}{\par\vspace{0ex}}
\makeatother

\let\orgautoref\autoref
\providecommand{\Autoref}[1]{\def\equationautorefname{Equation}
\def\figureautorefname{Figure}\def\sectionautorefname{Section}\def\subsectionautorefname{Section}\def\subsubsectionautorefname{Section}\def\tableautorefname{Table}\orgautoref{#1}}
\renewcommand{\autoref}[1]{\def\equationautorefname{Eq.}
\def\figureautorefname{Fig.}\def\sectionautorefname{Sect.}\def\subsectionautorefname{Sect.}\def\subsubsectionautorefname{Sect.}\def\tableautorefname{Table}\orgautoref{#1}}

\begin{document}

    \title{SHINE: Shear inference environment - Generative forward modelling for radio weak gravitational lensing}

\titlerunning{Shear inference for radio weak lensing}

%

   \author{
    Ezequiel Centofanti\inst{1}\fnmsep\thanks{E-mail: \href{mailto:ezequiel.centofanti@cea.fr}{ezequiel.centofanti@cea.fr}}\orcidlink{0009-0001-8461-8451}
    \and Benjamin Remy \inst{2,3}\orcidlink{0000-0002-0978-5612}
    \and Emma Ayçoberry \inst{1}\orcidlink{0000-0002-9235-1195}
    \and François Lanusse \inst{1,5}\orcidlink{0000-0001-7956-0542}
    \and Jean-Luc Starck\inst{1,4}\orcidlink{0000-0003-2177-7794}
    \and Samuel Farrens\inst{1}\orcidlink{0000-0002-9594-9387}
    }

   \institute{
    AIM, CEA, CNRS, Université Paris-Saclay, Université de Paris, F-91191 Gif-sur-Yvette, France
    \and NSF-Simons AI Institute for the Sky (SkAI), 172 E. Chestnut St., Chicago, IL 60611, USA
    \and Department of Astronomy and Astrophysics, University of Chicago, Chicago, IL 60637, USA
    \and Institutes of Computer Science and Astrophysics, Foundation for Research and Technology Hellas (FORTH), GR-70013 Heraklion, Greece 
    \and Flatiron Institute, New York, NY, USA
    }

   \date{Received Month XX, 20XX}
 
  \abstract{
    Next-generation radio interferometers, such as SKA, will observe the radio sky with unprecedented sensitivity and resolution. Their wide sky coverage will enable weak gravitational lensing studies using radio data, which not only complement optical observations but also probe higher redshifts. However, traditional shear estimation methods—based on measuring the ellipticity of observed galaxies—are not directly applicable to radio interferometric data.
    Interferometric observations are acquired in the Fourier domain, where galaxy images are delocalised, making shape measurements non-trivial. Furthermore, deconvolution methods such as CLEAN and its variants can introduce systematic errors when converting observations to the image domain, biasing shear estimates.
    To address these challenges, we propose a cosmic shear inference method based on forward generative modelling, which avoids explicit shape measurements. Our forward model simulates galaxy images, applies the shear transformation and, by including the instrumental response of the radio interferometer, simulates the visibilities of each galaxy individually, with the likelihood explicitly defined in Fourier space. Using Monte Carlo sampling algorithms, we estimate the posterior distribution of the cosmic shear conditioned on a set of galaxy observations that share a common shear.
    We obtain unbiased shear estimates with sub-percent accuracy when applied to parametric galaxy simulations. We compare our method to state-of-the-art shear estimators, demonstrating a reduction in uncertainty of up to $3$ times and no apparent multiplicative bias. 
    Furthermore, we carry out coverage tests to show that our posterior estimates are well calibrated.
    We then show that applying this parametric model to COSMOS galaxies with realistic morphologies introduces a bias in the shear estimate over $10\sigma$. 
    Finally, we demonstrate that, by incorporating a generative forward model based on an autoencoder architecture trained on real galaxy images, we can mitigate this bias to below the $2\sigma$ level.
  }


   \maketitle
   \nolinenumbers

\section{Introduction}

    The gravitational lensing effect describes how photons, emitted by distant sources, are deflected as they travel through space due to the presence of gravitational potentials between the sources and the observer, distorting the apparent shape and position of the observed objects \citep{bartelmann2010gravitational}. 
    This phenomenon provides a way to probe the large-scale structure (LSS) of the Universe by mapping the distribution of matter overdensity projected along the line of sight. 
    In the case of lensing by the LSS, the effect is very subtle, and the apparent deformation of the observed sources is minimal, with a change of less than $1\%$.
    This regime is known as weak gravitational lensing (WL)
    \citep{schneider2006WL, hoekstra2008WL, munshi2008cosmology, bartelmann2010gravitational, kilbinger2015, mandelbaum2018weak}, and the distortions can be modelled locally by a linear transformation that includes an isotropic magnification component, called convergence ($\kappa$), and an anisotropic stretching component, called shear ($\gamma$).
    The WL signal allows constraints to be placed on cosmological parameters related to the dark matter density and the growth of the LSS \citep{BARTELMANN2001291}. 
    In practice, the shear field is estimated statistically by measuring the shapes of large numbers of galaxies and averaging their observed ellipticities, under the assumption that their intrinsic orientations are randomly distributed. 
    However, given that the shear is very small compared to the ellipticity of the galaxies, it is necessary to measure the galaxy shapes with extreme precision.
    Furthermore, these measurements must be calibrated for the instrumental response of the telescope, the point spread function (PSF), which needs to be precisely modelled \citep{massey2012}.
    Finally, any additive or multiplicative bias in the shear estimate must be calibrated, requiring costly simulations, which highlights the need for unbiased methods. 
    
    The weak lensing signal has been successfully measured in past surveys using optical observations.
    Surveys such as CFHTLenS \citep{heymans2012_cfht}, KiDS \citep{KIDS_2013}, DES \citep{DES_2005}, and HSC \citep{miyazaki2012HSC} have made it possible to derive cosmological constraints from WL studies thanks to their depth, sky coverage, and compact PSF. 
    Upcoming optical WL surveys such as Euclid \citep{laureijs, euclid2024}, LSST \citep{LSST2009, Ivezi__2019}, or Nancy Grace Roman \citep{wfirst, akeson2019} will observe the sky with even greater depth and sensitivity, allowing for the detection of a higher galaxy density, and promising significant improvements in cosmological results.

    Radio WL is an evolving domain, and the detection of this signal using radio observations alone has not yet been conclusive. 
    Studies such as \citet{chang2004weak} report a detection in radio wavelengths, measuring the shapes of galaxies directly in Fourier space, but with a very low detection significance due to the low galaxy density detected. 
    Other studies such as \citet{tunbridge2016radio, demetroullas2018galaxy, hillier2019radio, liu2025prospects} perform a cross-correlation analysis, measuring galaxy shapes in both radio and optical observations. Overall, these studies highlight the major limitations of current radio telescopes in conducting WL studies, namely the low density of detected galaxies and the small survey areas compared to optical surveys \citep{patel2014weak,harrison2020superclass}.

    The next generation of radio telescopes such as the Square Kilometre Array (SKA) \citep{SKA_2009, SKA1_2019} and the Deep Synoptic Array (DSA) \citep{hallinan2019dsa} will observe the radio sky with unprecedented coverage (over $30\,000$ $\text{deg}^2$) and, thanks to their physical extension and sensitivity, will resolve a large number of galaxies, making WL radio studies possible \citep{SKA_2026, brown2015weak}. 
    Some of the most prominent advantages of this type of instrument is the ability to access to higher redshifts, or obtaining photometry directly from the $21\;\text{cm}$ HI line without relying on complementary observations \citep{blake2004cosmology}.
    Furthermore, polarised radio observations could allow the intrinsic shapes of galaxies to be estimated, alleviating key systematics in shear estimation, such as intrinsic alignments \citep{brown2011polarization}.
    Finally, another systematic that is of lesser relevance in radio is the PSF, commonly referred to as the dirty beam, which, unlike in optical telescopes, is well known and is determined by the spatial distribution of the radio antennas, the operating frequencies, and the direction of observation.  
    Conversely, radio surveys face other challenges for WL studies, such as a lower galaxy density and poorer resolution compared to optical telescopes.
    Moreover, the immense volume of data that will be acquired by the new generation of interferometers poses a major technical challenge for estimating cosmic shear, requiring highly efficient and scalable algorithms.

    This paper is organised as follows: in \autoref{sec:SOTA} we discuss state-of-the-art methods for shear estimation in radio WL, in \autoref{sec:theory} we introduce the theoretical background, particularly the WL formalism and radio interferometry imaging, in \autoref{sec:method} we present our forward modelling framework for shear estimation, In \autoref{sec:data} we explain the simulated dataset used in this work, in \autoref{sec:results} we present the experimental results and finally describe future prospects and conclude in \autoref{sec:conclusions}. 

\section{Related work}
\label{sec:SOTA}
    The traditional procedure for estimating cosmic shear consists of measuring the shapes of galaxies individually. 
    There are two main ways of doing this: one approach is to measure galaxy ellipticity directly from images, for example by computing second-order moments, deriving these moments from a decomposition of the image into basis functions, or using convolutional neural networks. 
    The other approach consists of assuming a model of the galaxy surface brightness and ellipticity, and either individually or collectively fitting them to the observations. For a review of shape measurement methods see \citet{mandelbaum2015great3}.
    The advantage of the second approach is that the PSF can be included in the forward model, whereas the first approach requires separate PSF calibration. 
    However, the forward model approach is susceptible to model bias if it fails to represent the morphology of the observed galaxies.
    Model bias will require further calibration of the estimated shear, adding computational overhead. This calibration cost may be limiting for large surveys such as \textit{Euclid}, Nancy Grace Roman Space Telescope, and SKA, which will observe trillions of galaxies.

    Radio astronomy poses another challenge, which is the fact that observations are not acquired directly in the image domain but in Fourier space, where the definition of galaxy shape and ellipticity are not trivial. 
    One option is to reconstruct the galaxy image from the Fourier samples (commonly called $uv$ samples or visibilities in radio interferometry) using a deconvolution algorithm such as CLEAN \citep{hogbom1974aperture} or its more recent versions \citep{cornwell2008multi, rau2011multi, garsden2015lofar, connor2022deep, dia2025iris} which could introduce artefacts in the shape of the reconstructed galaxy due to the complex structure of the radio PSF.
    In \citet{patel2014weak}, real-space shapelets are applied to CLEAN-reconstructed images to measure the shapes of galaxies.
    The other option is to measure the shape directly in Fourier space. In \citet{chang2004weak} a method based on shapelets, introduced in \citet{chang2002shape}, is used to estimate the shape of galaxies in Fourier space directly from interferometric data.
    Other works such as \citet{tripathi2025deepshape} use convolutional neural networks to extract features from both the galaxy image and the radio dirty beam to estimate galaxy ellipticities. However, this type of method will require retraining when applied to real data, for which there are no labels for the true shape of the observed galaxies or deconvolved images. 
    Frameworks such as \citet{rivi2016radio} model the light profile of galaxies analytically in $uv$ space, which allows the likelihood to be evaluated in Fourier space directly using visibilities. However, their analytical model is limited to exponential profiles, which are overly simple for well-resolved galaxies.
    Lastly, the approaches mentioned above have a potential source of model bias in the shear estimation due to the simplification of galaxy shapes to the ellipticity, which is not a well defined quantity for arbitrary galaxy morphology \citep{remy2022towards}. 

    In this paper, we present a method for estimating shear from radio galaxies that it is free from shape measurements. Our framework models shear hierarchically for multiple galaxies within the same region of the sky. This method consists of a forward galaxy model that includes both the shear distortion and the instrumental response of the radio interferometer, simulating the visibilities of each galaxy individually.
    We leverage two different galaxy generative models, adapting to the morphological complexity of the observed galaxies in order to quantify and address potential model biases in the case of a morphological mismatch. 
    

    \begin{figure*}
        \centering
        \includegraphics[width=.6\linewidth,trim={0.0cm 0.0cm 0.0cm 0.0cm},clip]{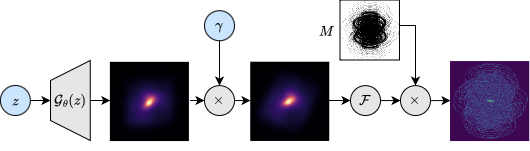}
        \caption{Diagram of the generative forward model. $\mathcal{G}_\theta(z)$ is a galaxy image generative model. The morphology of galaxies is encoded in the vector $z$, which may consist of real parameters ($hlr$, flux, ellipticity, etc.) or a latent representation of the images. The shear operation, $\cdot\times\gamma$, is then applied, and finally, the sheared galaxy is sampled in Fourier space by the $uv$ mask $M$ to simulate the visibilities.}
        \label{fig:forward_model}
    \end{figure*}

\section{Theoretical background}
\label{sec:theory}

    \subsection{Weak gravitational lensing}

    The non-homogeneous distribution of dark matter between distant galaxies and the observer slightly deflects the emitted photons, modifying the apparent shape of the observed galaxies. In the case of lensing due to the LSS, this deformation is subtle and can be modelled locally by a linear transformation described by the following matrix: 
    \begin{equation}
        A = \begin{bmatrix} 1 - \kappa - \gamma_1 & - \gamma_2
    \\ - \gamma_2 & 1 - \kappa + \gamma_1 \end{bmatrix},
    \end{equation}
    where $\kappa$ is a scalar representing an isotropic magnification of the observed shape, and $\gamma=\gamma_1+i\gamma_2$ is a complex number that models an anisotropic stretching. 
    These two quantities are closely related since they are given by second-order derivatives of the two-dimensional lensing potential, $A_{ij}=\delta_{ij}-\partial_i \partial_j \psi$ (see \citet{bartelmann2010gravitational, kilbinger2015}).
    Thus, a source with intrinsic ellipticity $\epsilon^{\text{i}}$, when distorted by the WL phenomenon, will be observed with the following ellipticity 
    \begin{equation}
        \epsilon^{\text{obs}}=\epsilon^{\text{i}}+\gamma.
    \end{equation}
    Note that we are not considering the size change by $(1-\kappa)$, as in the context of WL these distortions are very small and we can approximate the reduced shear as follows
    \begin{equation}
        g = \frac{\gamma}{1-\kappa} \approx \gamma.
    \end{equation}

    One way to estimate shear is by averaging the observed ellipticities of multiple sources in the same region of the sky, where the shape distortion is assumed to be the same for all objects, obtaining
    \begin{equation}
        \langle \epsilon^{\text{obs}} \rangle = \langle \epsilon^{\text{i}} \rangle + \gamma.
    \end{equation}
    Then, since the intrinsic shapes of galaxies do not have a preferred orientation, the average of intrinsic ellipticities cancels out, yielding   
    \begin{equation}
        \langle \epsilon^{\text{obs}} \rangle = \gamma + n,
    \end{equation}
    where n represents shape noise, which depends on the number of averaged galaxies. In this work, we do not model the possible intrinsic alignment (IA) of background galaxies with their hosting LSS, which can correlate galaxy shapes even in the absence of lensing \citep{schafer2009galactic}. 
    
    Estimating the reduced shear from noisy ellipticity measurements yields a biased estimate, because the ellipticity estimator responds non--linearly to noise and to imperfect galaxy and PSF models. To first order this bias is described by a linear model,
    \begin{equation}
        \langle \epsilon^{\text{obs}} \rangle = (1+m)\,g + c,
    \end{equation}
    with a multiplicative bias $m$ and an additive bias $c$. Calibration methods such as metacalibration \citep{huff_2017_metacal, Sheldon_2017_metacal, Sheldon_2020} characterise the response to shear directly from the data and correct the estimate, at the cost of additional shape measurements for each galaxy.

    \subsection{Radio interferometric imaging}
    Radio interferometers, unlike single-dish radio telescopes, use multiple radio antennas to perform observations. 
    The main advantage is that the angular resolution of the telescope depends on the maximum separation between antennas instead of the size of the individual dish reflectors. 
    This allows resolutions comparable to optical telescopes ($\text{arcsec}$) to be achieved without the need to build extremely large, expensive antennas that are challenging to deploy and maintain.
    Radio interferometers correlate the incoming radio signals between pairs of antennas. 
    This allows, through the interferometry principle, to decompose the observed sky into sinusoidal components. 
    The complex coefficients associated with these components are called visibilities, and their relationship with the observed sky is given by the van Cittert-Zernike theorem \citep{Thompson_interferometry}.
    \begin{equation}
    \label{eq:vCZ}
        \mathcal{V}(u,v,w) = \iint\frac{i_{{\text{sky} }}(l,m)}{n} e^{-2\pi j(ul+vm+w(n-1))}{\text{d}}l{\text{d}}m,
    \end{equation}
    where $\mathcal{V}(u,v,w)$ is the complex visibility function, $i_{{\rm sky}}$ is the observed sky intensity as a function of the two direction cosines $(l, m)$ and $n$ is the third direction cosine
    \begin{equation}
        n = \sqrt{1-l^2-m^2}.
    \end{equation}
    If we assume that the observed sources do not subtend large angles in the sky, the small sky approximation is fulfilled
    \begin{equation}
        l^2+m^2 \rightarrow 0,
    \end{equation}
    which simplifies \autoref{eq:vCZ} yielding
    \begin{equation}
    \label{eq:vCZ_2d}
        \mathcal{V}(u,v) = \iint i_{{\rm sky}}(l,m) e^{-2\pi j(ul+vm)}{\rm d}l{\rm d}m.
    \end{equation}
    Neglecting the $w$ term introduces a distortion associated with projecting the celestial sphere onto a plane \citep{taylor1999imag}. Nevertheless, if the imaged region is sufficiently small and centred on the direction of observation, this effect can be ignored \citep{cornwell2008wproj}.
    We can rewrite \autoref{eq:vCZ_2d} in terms of the two-dimensional (continuous) Fourier transform
    \begin{equation}
        \mathcal{V}(u,v) = \mathcal{F}\left[ i_{{\rm sky}}(l,m) \right](u,v).
    \end{equation}
    However, since each pair of antennas samples the complex visibility function at a specific position in the $uv$ plane, which depends on their separation, known as the baseline, only a limited number of noisy $uv$ points are measured 
    \begin{equation}
       \mathcal{V}_{i} = I_{{\rm sky}}(u_i, v_i) \;+\; N(u_i,v_i) .
    \end{equation}
    If we represent the set of $uv$ sampling positions by a mask in Fourier space, we can describe the measured visibilities as follows:
    \begin{equation}
    \label{eq:vis}
        V = M \;\odot\; \left(F \,i_{{\rm sky}} + N\right),
    \end{equation}
    where $V$ are the measured visibilities, $F$ is the discrete Fourier transform, $M$ is the sampling mask (containing ones at $uv$ positions with samples and zeros elsewhere), $\odot$ is the Hadamard product which represents the element-wise matrix multiplication, and $N$ is the additive Gaussian noise. Note that the noise, which is white in $uv$ space, is correlated through the sampling mask in real space.

\section{Methodology}
\label{sec:method}

    The proposed approach to estimate cosmic shear consists of forward modelling the visibilities of the observed galaxies rather than performing individual shape measurements on the galaxy images. 
    The forward model allows us to evaluate the likelihood of the observed visibilities conditioned on the galaxy morphology parameters and the shear.
    Using Monte Carlo sampling methods, we are able to approximate the full posterior shear distribution. 
    This approach aims to obtain an unbiased estimate of the shear without the need to measure galaxy shapes explicitly.

    \subsection{Generative forward model}
        The forward model consists of a generative galaxy model $\mathcal{G}_{\theta}(z)$, a pixel-level shear transformation, and a model of the instrumental response of the radio interferometer. \Autoref{fig:forward_model} illustrates the forward model and provides examples of its intermediate outputs.
        This model is coded in JAX \citep{jax2018github}, a library for array-oriented numerical computations, with automatic differentiation and GPU compatibility and uses JAX-GalSim, a differentiable, GPU accelerated and scalable implementation of the GalSim \citep{rowe2015galsim} library, for modelling the shear operation. This facilitates the use of sampling methods such as Hamiltonian Monte Carlo (HMC) or Microcanonical Langevin Monte Carlo (MCLMC), which require calculating gradients of the likelihood with respect to the sampled parameters.
    
        \subsubsection{Galaxy generative model}
        \label{sec:generative_model}
            The generative model $\mathcal{G}_\theta$ takes as input a set of galaxy morphology parameters, z, and produces an image of the associated galaxy in pixel space. 
            We consider two types of generative models, each adapted to different morphological complexities.

\begin{figure}
                \centering
                \includegraphics[width=1.\linewidth]{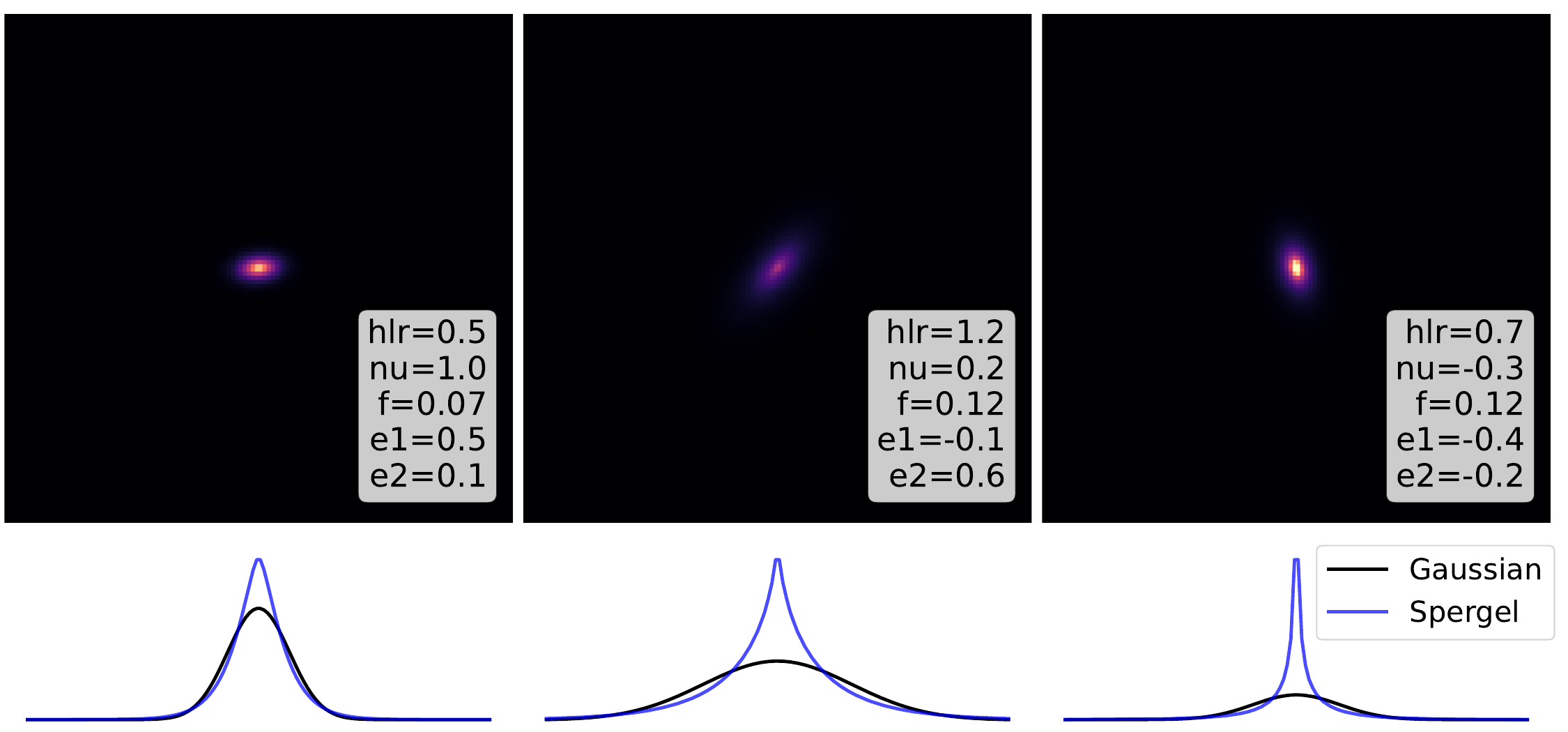}
                \caption{Three Spergel profiles with their respective galaxy parameters. The three images are plotted on the same scale. A zoomed 1D slice is shown below each profile together with a Gaussian profile of the same half light radius and integrated flux.}
                \label{fig:spergel_profiles}
            \end{figure}

            \paragraph{Parametric generative model}
            This model represents the surface brightness profile of galaxies using a Spergel parametric profile \citep{spergel2010analytical}.
            The advantage of this profile is that it has an analytical Fourier transform, making it ideal for modelling visibilities in $uv$ space.
            We use the implementation of this profile available in JAX-GalSim. In this model the galaxy parameters ($z$) that define the output image are: the half-light radius ($hlr$), which is related to the apparent size of the galaxy, the flux ($f$), which is related to the amplitude of the surface brightness profile, the Spergel index ($\nu$), which controls the degree of central concentration of the profile, and the intrinsic ellipticity ($e_1, e_2$). In this case, the generative model does not have any parameter $\theta$ that need to be learned or tuned other than the parameter priors, that will be discussed later. \Autoref{fig:spergel_profiles} shows some examples of Spergel profiles in real space prior to being convolved with the radio PSF.

\begin{figure}
                \centering
                \includegraphics[width=.9\linewidth, trim={1.3cm 0cm 1.3cm 0cm},clip]{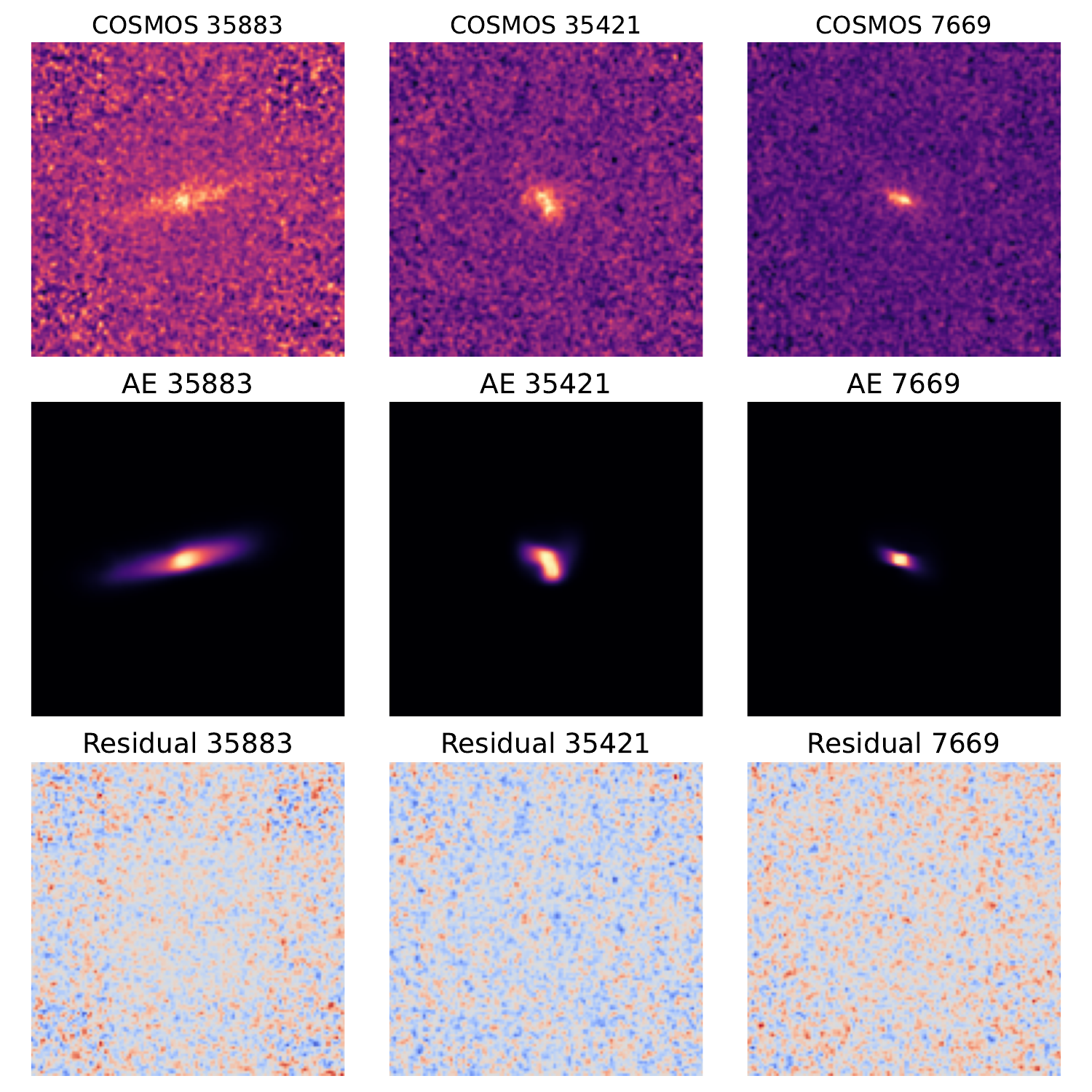}
                \caption{COSMOS real galaxies examples, their corresponding AE uncovolved surface brightness profile reconstruction, and the respective residuals. }
                \label{fig:ae_profiles}
            \end{figure}

            \paragraph{Deep NN generative model}
            The second generative model is based on a Auto-Encoder (AE) neural network (NN) architecture, similar to that proposed in \cite{lanusse2021deep}.
            We use a Deep Compression Auto Encoder (DC-AE) model with residual autoencoding layers, introduced in \citet{chen2024deep}.
            The AE architecture has two components: an encoder that compresses an input image into a much smaller vector known as the latent representation of the input image. And the second component is a decoder, which is trained to take the latent representation and reconstruct back the original image. This two components have a mirrored architecture and use multiple convolutional layers to compress and reconstruct back the input image. 
            For an overview of recent autoencoder models, see \citet{bank2023autoencoders}.
            In our case, we use an encoder with 5 hidden layers that use residual convolutional networks \citep[ResNet, ][]{he2016deep} and 2D convolutional layers that reduce the size of the input image ($128\times128)$ to a latent variable of ($4\times4$). The decoder has an equivalent structure but with pixel-shuffle up-convolution layers \citep{shi2016real} to increase the dimensions of the latent representation to match the size of the original image.
            The AE is coded in \texttt{equinox} \citep{kidger2021equinox}, a JAX library for neural networks, which allows for direct integration into the forward model.
            The AE architecture and the training script are available on GitHub\footnote{\url{https://github.com/b-remy/probabilistic-shear}}. 
            Given the lack of observations of well-resolved galaxies in the radio band, we used optical observations. 
            The AE is trained using $80\,000$ images of real galaxies with complex morphologies taken from the COSMOS \citep{scoville2007cosmos} survey, optical observations from the Hubble Space Telescope (HST). 
            The dataset is split into $70\,000$ training examples and $10\,000$ test examples, and we incorporate random flipping of the galaxy images and their PSF to augment the data.
            Convolution with the HST PSF is applied at the decoder output so that the AE learns the unconvolved surface brightness profile of the COSMOS galaxies.
            The loss function used for training the AE is the mean absolute error (MAE) with total variation (TV) regularisation, and is written as follows
            \begin{equation}
                \mathcal{L}_{\theta}(\hat{x}) = \|x-H\ast\hat{x}\|_1 + \lambda (\|\nabla_1\hat{x}\|_1 + \|\nabla_2\hat{x}\|_1),
            \end{equation}
            where $x$ is the real COSMOS galaxy image, $\hat{x}$ is the AE unconvolved galaxy reconstruction, $H$ is the corresponding HST PSF and $\nabla_{i}$ is the gradient with respect to the $i$-th coordinate, computed as the difference of adjacent pixels.
            The first term corresponds to the data fidelity, and the second term,  weighted by $\lambda=1e^{-5}$, is the total variation, which drives the denoising and smoothness of the unconvolved profiles.
            The classic Kullback–Leibler (KL) divergence penalty on $z$, used to enforce a Gaussian distribution in the latent space, is not included since it tends to degrade the quality of the images generated by the autoencoder.
            Instead, a softplus-type saturation, $z'=z/\sqrt{1+{z^2}/{25} }$, is used, matching the range of a Gaussian distribution, as suggested in \citet{rozet2026lost}.
            Once the AE is trained the decoder part can be used as a generative galaxy model, which is much more expressive than parametric profiles and can generate galaxies with well-resolved features. In this model, the galaxy parameters are replaced by the latent representation $z\in \mathbb{R}^{4\times4}$. This allows us to generate galaxy images for different values of z. 
            \Autoref{fig:ae_profiles} shows some examples of real COSMOS galaxies alongside the AE reconstruction, prior to convolution with the HST PSF and the respective residuals.

            \paragraph{Normalising flow}

            To generate new images of galaxies, we must sample new latent vector values. However, since no KL regularisation is used in the latent space, the distribution $p(z|X_{\rm obs})$ is unknown, making the sampling of new latent vectors a non-trivial task. 
            In order to be able to draw samples in the latent space we use a normalising flow, which allows us to learn a bidirectional mapping between the target distribution $p(z|X_{\rm obs})$ and a base distribution, typically a standard Gaussian $u \sim \mathcal{N}(0,I)$. For a review of normalising flows, see \cite{papamakarios2021normalizing}.  
            Since the flow is used in the context of differentiable probabilistic sampling-which requires forward modelling of several thousand galaxies and the calculation of gradients with respect to galaxy parameters—we need a flow that is fast in the forward direction and provides efficient gradient evaluation. 
            The Real NVP flow \citep{dinh2016density} is a type of coupling flow that exhibits computational symmetry and allows for efficient evaluation.
            The coupling layers split the latent vector in two (usually in half, i.e. $d = D/2$) and apply a bijection of the form
            \begin{align}
                u'_{0:d} &= u_{0:d} \nonumber \\
                u'_{d+1:D} &= u_{d+1:D} \odot \exp(s(u_{0:d})) + t(u_{0:d}),
            \end{align}
            where $u$ and $u'$ are the input and output of the coupling layer, $s$ and $t$ are scaling and translation functions, and $\odot$ is the Hadamard or element-wise product. 
            The functions $s$ and $t$ consist of multilayer perceptron (MLP) neural networks whose parameters are learned during the training of the flow. 
            In this work, we use a Real NVP flow comprising four coupling layers, with a two-layer MLP of $128$ neurons and ReLU activation.
            We use the FlowJAX coupling flow implementation \citep{ward2025flowjax}.
            The flow is trained by minimising the KL between the true prior distribution $p(z|X_{\rm obs})$ and the variational distribution $q(z|X_{\rm obs})$.
            In practice, this is achieved by minimising the negative log-likelihood over the variational distribution.
            We train the flow on $80\,000$ latent representations of the COSMOS dataset obtained with the encoder model of the trained AE. The flow is incorporated into the forward model right before the decoder, thus, instead of sampling the latent space $z$, we sample from a Gaussian distribution $u \sim \mathcal{N}(0,1)$ and obtain $z=f(u)$ using the flow.
            
        \subsubsection{Instrumental response}
        \label{sec:radio_psf}
\begin{figure}
                \centering
                \includegraphics[width=0.8\linewidth, trim={.7cm 0cm 2.cm 1.45cm},clip]{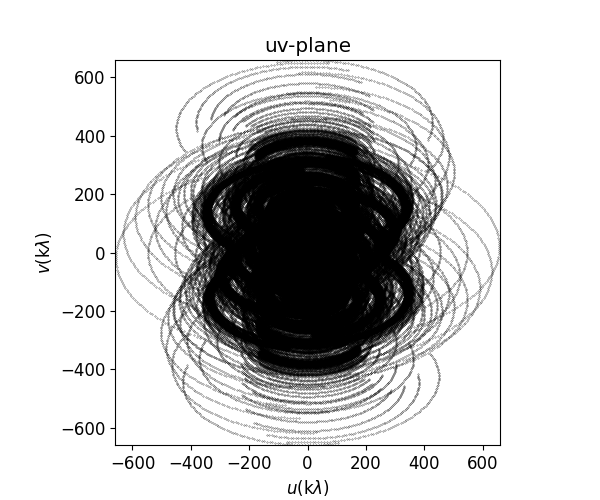}
                \caption{$uv$ coverage for the SKA-Mid antenna layout.}
                \label{fig:uv_mask}
            \end{figure}

            Once the galaxy image has been generated and shearing is applied using JAX-GalSim, the instrumental response is applied in Fourier space.
            The radio PSF is modelled as described in \autoref{eq:vis}, using a sampling mask ($M$) in $uv$ space. This set of samples constitutes the simulated visibilities.
            We simulate the $uv$ sampling mask with \texttt{argosim} \citep{centofanti2026argosim}\footnote{\url{https://github.com/ARGOS-telescope/argosim}}, a modular, lightweight, and JAX-based package for radio interferometric simulations. 
            We use SKA1-Mid AA4 as our reference interferometer. We calculate the $uv$ tracks for an 8-hour observation, sampling every 5 minutes. 
            We simulate an observation at zenith, which corresponds to a declination of approximately $-30$ degrees.
            We simulate the $uv$ tracks for a single frequency band centred at $1.4$ GHz.
            The resulting $uv$ coverage is shown in \autoref{fig:uv_mask}.
            In this work, we simulate a single galaxy per FOV and assume that all the simulated galaxies are located at the same pointing centre, using the same radio PSF for all simulated galaxies. This should not affect the results, as the $uv$ mask is deterministic, but it does greatly simplify data generation.
            In reality, effects such as radio frequency interference (RFI) or calibration errors can alter the radio PSF, making it not fully deterministic. However, these details are not addressed in this work.
            The $uv$ samples are gridded to $128 \times 128$ pixels to match the size of the simulated galaxy images, using a pixel scale of $0.15\;\rm{arcsec}\;\rm{pixel}^{-1}$.
            At this pixel scale, the $uv$ samples associated with the longest baseline fit just within the $128 \times 128$ image.

    \subsection{Posterior inference via differentiable sampling}
        Forward modelling the visibilities allows us to model the likelihood of the observed data.
        The model for the observation of a single galaxy is written as
        \begin{equation}
            V_{\text{obs}} = M \odot F(\gamma \otimes \mathcal{G}) + N,
        \end{equation}
        where $M$ is the $uv$ sampling mask, $F$ is the Fourier transform, $\gamma\otimes\cdot$ represents the shear operation and $N$ is noise added to the sampled visibilities. 
        The noise in $uv$ space is Gaussian and independent per visibility.
        Then, for the observation of $N_{\rm gal}$ galaxies, the likelihood is written as follows
        \begin{equation}
            p(V_{\text{obs}}|z,\gamma) \propto  \exp \left( - \sum_{i=1}^{N_{\text{gal}}} \sum_{j=1}^{N_{\text{vis}}}\frac{|V_{\text{obs\_i}}[j] - V(z_i,\gamma)[j]|^2}{\sigma_j^2} \right),
        \end{equation}
        where $V_{\text{obs\_i}}[j]$ is the $j$-th visibility sample of the $i$-th observed galaxy, and $\sigma_j$ is the noise associated with the $j$-th baseline. 
        Finally, we can write the joint posterior for the galaxy parameters and the shear in terms of the likelihood using Bayes’ theorem, resulting in
        \begin{equation}
            p(z, \gamma | V_{\text{obs}}) \propto p(V_{\text{obs}}|z,\gamma)p(z)p(\gamma).
        \end{equation}
        Since the galaxy morphology parameters and shear are uncorrelated, we can factorise the prior as the product of two independent priors, $p(z,\gamma) = p(z)p(\gamma)$. 
        
       \subsubsection{Monte Carlo sampling}
       \label{sec:sampling}
            To estimate the posterior distribution, we use Monte Carlo sampling methods.
            Given that we are dealing with multiple galaxies simultaneously, the number of morphological parameters is considerably large (over $700$ parameters for $N_{\rm gal}=100$), making the use of conventional sampling methods such as Metropolis-Hastings unfeasible.
            We therefore use Microcanonical Langevin Monte Carlo (MCLMC), a gradient-based sampling method that is highly efficient for sampling in high-dimensional spaces \citep{robnik2023mclmc}. MCLMC automatically tunes its hyperparameters and is orders of magnitude faster than other gradient-based methods such as Hamiltonian Monte Carlo \citep[HMC,][]{betancourt2017hmc}.
            These algorithms model the trajectory of the sampled particles using dynamics inspired by physical systems.
            In particular, they use Hamiltonian dynamics, which utilise the gradients of the target distribution to simulate the trajectories of the sampled particles.
            We exploit the fact that our forward model is built in JAX meaning that the likelihood is fully differentiable with respect to the sampled parameters. 
            We use the MCLMC implementation provided by the \texttt{blackjax} Bayesian inference library \citep{cabezas2024blackjax}.
            To increase the number of independent samples, we run multiple chains in parallel. All chains are initialised randomly from points drawn from the prior distributions of the tracked parameters. 
            Then, before running the MCLMC adaptation and sampling, we perform a maximum a posteriori (MAP) estimation for each chain using gradient descent.
            After sampling, we verified the convergence of the chains using the Effective Sample Size (ESS) statistic, which measures the number of uncorrelated samples for each of the tracked parameters.  

        \subsubsection{Kernel density estimation}
            Once the Monte Carlo chains have taken a sufficient number of independent samples from the posterior distribution, we proceed to estimate the probability density. 
            We used a simple Gaussian mixture model (GMM) after verifying that the distribution of the posterior samples is fairly round and unimodal. 
            The posterior density is then written as follows
            \begin{equation}
                p(\gamma|V_{\rm obs}) \sim \sum_{k=1}^{K}w_k\,\mathcal{N}(\mu_k,\,\Sigma_k),
            \end{equation}
            with $K$ Gaussian components.
            We use the \texttt{scikit-learn} ML library \citep{scikit-learn} to fit the GMM parameters, fixing the number of components at $K=5$.
            To combine multiple posterior distributions, in cases where the posterior has a single mode and a Gaussian-like shape, we perform the analytical multiplication of the likelihoods, dividing $N-1$ times with the prior ($\Sigma_p)$ , yielding 
            \begin{equation}
                \Sigma_c =\left( \Sigma_p^{-1}+\sum_{i=1}^N \left(\Sigma_i^{-1}-\Sigma_p^{-1}\right)\right)^{-1}, \;\;\;\: 
                \mu_c = \Sigma_c \left( \sum_{i=1}^N \Sigma_i^{-1}\mu_i \right).  
            \end{equation}

\section{Radio simulations}
\label{sec:data}

    Given the absence of radio observations with sufficient resolution and galaxy density to perform a WL-scale analysis, in this study we apply the proposed methodology to simulated data.
    Furthermore, this allows us to evaluate the performance of our method in a more controlled setting, given that we know the ground truth value of the applied shear. 
    We simulated two datasets of galaxy observations with varying degrees of morphological complexity. 
    In the first dataset each galaxy is represented by a single component parametric profile, and in the second dataset we turn real optical images of galaxies from the COSMOS survey \citep{scoville2007cosmos} into radio observations.

    \subsection{Parametric radio galaxies}
    \label{sec:data_param}

        The parametric radio galaxies are modelled by Spergel profiles using JAX-GalSim. Spergel profiles can be described analytically both in real space and Fourier space, and offer greater flexibility than exponential or Gaussian profiles.  
        These profiles are characterized by two main parameters: their size, represented by the half-light radius (hlr), and the Spergel index $\nu$.
        The form of the Spergel profile in the 2D-projected Fourier space is as follows
        \begin{equation}
            \Sigma_\nu(k) = \frac{L_0}{2\pi \left[1 + (k_x^2+k_y^2)\left(\frac{r_{\rm hlr}}{c_\nu}\right)^2 \right]^{1+\nu}},
        \end{equation}
        where $c_\nu$ is a tabulated constant that depends on $\nu > -1$. In real space, this profile is described as follows
        \begin{equation}
            \Sigma_\nu(r) = \frac{c_\nu^2 L_0}{r_{\rm hlr}^2} f_\nu\left( \frac{c_\nu r}{r_{\rm hlr}} \right),
        \end{equation}
        where $f_\nu(u) = (u/2)^\nu K_\nu(u)/\Gamma(\nu+1)$ and $K_\nu(u)$ is a modified spherical Bessel function of the third kind.
        The analytical expressions of this profile are described in more detail in \cite{spergel2010analytical}.
        The Spergel profile can approximate an exponential profile for $\nu=0.5$, equivalent to a Sérsic index $n=1$, and a de Vaucouleurs-type profile for $\nu=-0.6$, equivalent to a Sérsic index $n=4$.
        In \cite{tan2024fitting} the authors present a comparison between the Sérsic and Spergel profiles and provide an analysis of the ranges of values for the Sérsic index $n$ and the Spergel index $\nu$ that result in similar surface brightness.
        We believe that, for this study, the type of profile used is not a determining factor, as they share a similar morphological complexity.
        An important factor in our preference for the Spergel profile over the Sérsic profile is that the former has a differentiable JAX-based implementation in JAX-GalSim, which is essential for constructing a fully differentiable forward model. 

        \begin{figure}
                \centering
                \includegraphics[width=.75\linewidth,trim={0.cm 1.2cm 3cm 2.7cm},clip]{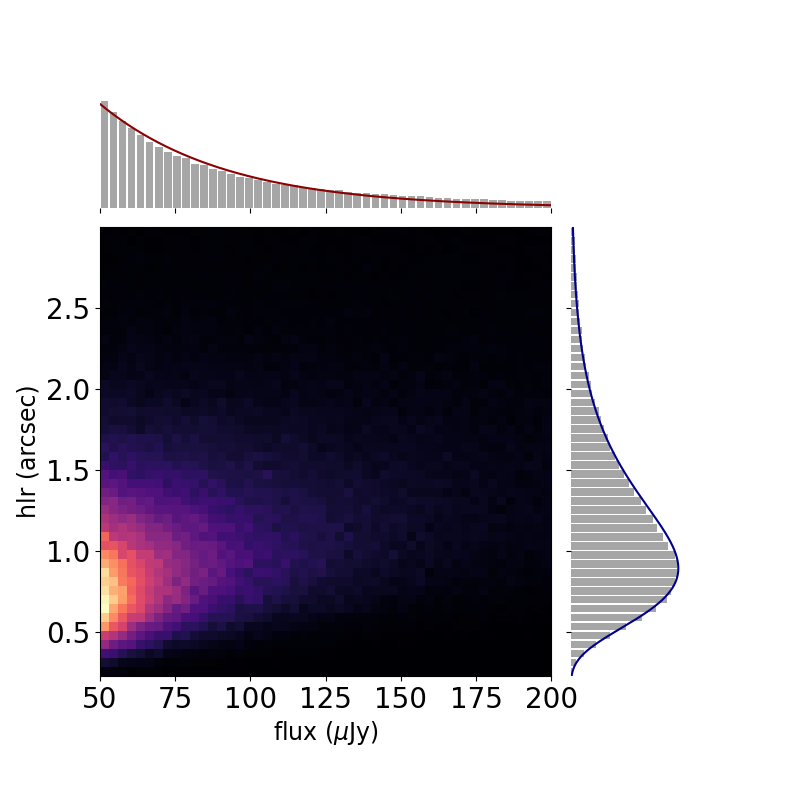}
                \caption{Histogram of the joint distribution of hlr and flux parameters for the selected T-RECS sources. The marginal histograms are shown at the top and on the left, alongside the respective density fits. }
                \label{fig:trecs_params}
        \end{figure}
        
        Once the Spergel profiles have been generated, we assign an intrinsic ellipticity and flux to each individual galaxy, and apply the same fiducial shear to all the galaxies. In this study, we tested various shear values, which we discuss later in \autoref{sec:results}.
        For the population of the parametric dataset, we used the Tiered-Radio Extragalactic Continuum Simulation (T-RECS) presented in \cite{TRECS}.
        These simulations provide a catalogue of star-forming galaxies that matches the flux, size, shape, redshift, among other sources parameters, to recent observations.
        As in \cite{tripathi2025deepshape} we used a selection cut in the integrated flux, ranging from $50$~{\textmu}Jy to $200$~{\textmu}Jy.
        \Autoref{fig:trecs_params} shows the joint distribution of flux and size for the selected galaxies. We randomly sample independent flux and hlr values from exponential and log-normal distributions fitted to the T-RECS catalogue. The exponential and log-normal fits are shown in red and blue, respectively, over the marginal histograms in \autoref{fig:trecs_params}. 
        With regard to the Spergel index $\nu$, it is sampled from a logistic-normal\footnote{$y \sim {\rm LogisticNormal} (a,b)$, defined as $y=a+(b-a)\,{\rm sigmoid}(x)$, where $x\sim\mathcal{N}(0, I)$.} distribution (not to be confused with log-normal) scaled to $\nu_{\rm max}=1$ and $\nu_{\rm min}=-0.7$, limiting the Spergel index to its valid range of application.
        Finally, the intrinsic galaxy ellipticities are sampled form independent distributions $\rm e_i \sim \mathcal{N}(0, \sigma_{\rm e})$ with $\sigma_{\rm e}=0.2$. To ensure the numerical stability of the Galsim shear operation, we limit the ellipticity to a unit modulus. 

        The simulated galaxies are drawn in Fourier space, where the $uv$ mask described in \autoref{sec:radio_psf} is applied to obtain the visibilities. Galaxies are simulated separately, each yielding its own $uv$ set. Next, we add independent noise to each visibility to obtain the noisy observations. The signal-to-noise ratio (S/N) is defined as follows
        \begin{equation}
            {\rm S/N} = \sqrt{\sum_{j=1}^{N_{\rm vis}}\frac{|V[j]|^2}{{\sigma_{uv}}_j^2}}.
        \end{equation}
        The noise level is adjusted to achieve a S/N ranging from 10 to 80.
        In this study, we use the same noise level for all the visibilities. The noise level used for the parametric galaxies is $\sigma_{uv}=0.02$.
        
    \begin{figure}
            \centering
            \includegraphics[width=\linewidth, trim={0cm .0cm 0.cm -.3cm},clip]{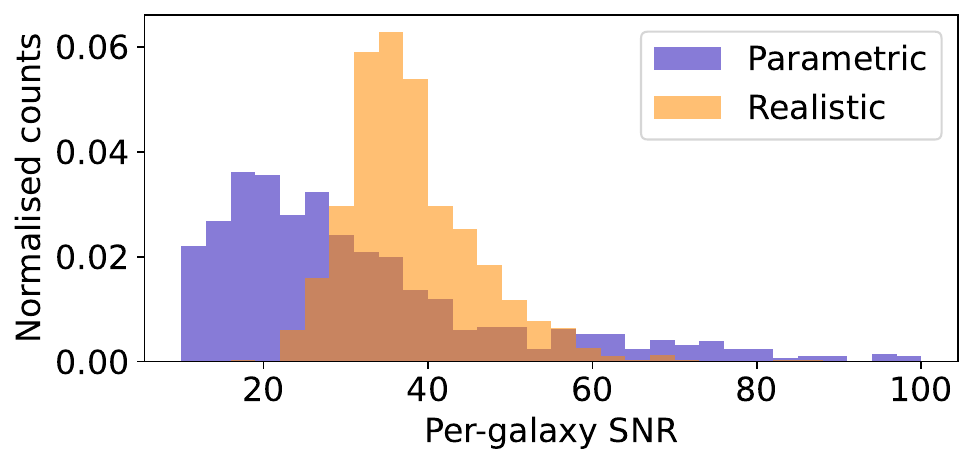}
            \caption{S/N distributions for the parametric and realistic simulated galaxies.}
            \label{fig:snr_distributions}
    \end{figure}
    
    \subsection{Realistic radio-like galaxies}
    \label{sec:data_cosmos}
\begin{figure}
            \centering
            \includegraphics[width=\linewidth, trim={.4cm .0cm 0.8cm 0cm},clip]{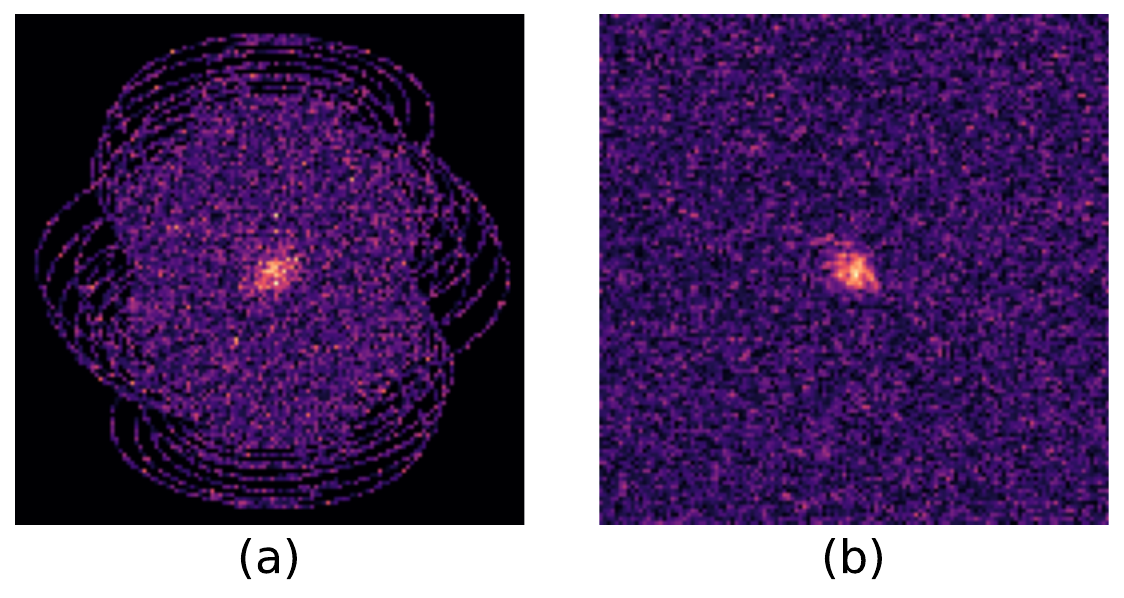}
            \caption{Whitened real COSMOS galaxy adapted to radio. (a) Noisy visibilities (${\rm S/N}=30$). (b) Galaxy dirty image.}
            \label{fig:cosmos_radio_like}
    \end{figure}

    We constructed a second dataset containing more complex galaxy morphologies using observations of real galaxies from the COSMOS survey.
    The COSMOS galaxies consist of observations taken by the Hubble Space Telescope (HST) in the optical band.
    Given that optical telescopes have higher resolution and sensitivity than radio telescopes, they are able to detect a greater number of well-resolved galaxies. 
    In our efforts to consider complex galaxy morphologies, we artificially adapted these galaxies to radio-like observations. 
    In \citet{patel2010radio}, the authors analyse the correlation between the shapes of galaxies observed using both optical telescopes (HST) and radio interferometers (VLA). They conclude that, on a galaxy-by-galaxy basis, optical and radio shapes are weakly correlated; however, the shape distributions in the optical and radio sky were very similar.
    We use the COSMOS 25.2 catalogue available on GalSim, which contains approximately $87\,000$ galaxies with the corresponding original HST PSF. 
    We applied a magnitude cut at $24.0$, excluding large galaxies with highly resolved features, resulting in a sample of approximately $40\,000$ galaxies.
    The COSMOS galaxies are co-added images, that is, they are a combination or stack of multiple observations, which results in a non-trivial noise distribution.
    Furthermore, as we need to construct a mock dataset of sheared galaxies and it is not preferable to shear the noisy images, we whitened\footnote{The galaxies were denoised using the AE and pure (white) Gaussian noise was added after the shear operation.} the COSMOS galaxies.
    We used the trained AE described in \autoref{sec:generative_model} to reconstruct the unconvolved surface brightness profile of the COSMOS galaxies.
    Then, we applied a fiducial shear to the whitened galaxies and drew them in Fourier space with a size of $128\times128$ pixels, using the pixel scale of the COSMOS images ($0.03 \; \rm{arcsec\;pixel}^{-1}$).
    Finally, we obtain the complex visibilities by applying the radio PSF in Fourier space as described in \autoref{sec:radio_psf}.
    In this dataset, we use two different pixel scales: one for generating the COSMOS galaxies and another for the radio PSF ($0.15 \; \rm{arcsec\;pixel}^{-1}$). 
    This artificially shrinks the radio PSF by a factor of $5$, so that the radio adapted COSMOS galaxies are well resolved. This is because the simulated radio PSF is much larger than the HST PSF.
    To model observational noise, we add independent Gaussian noise to the $uv$ samples as described in \autoref{sec:data_param}.
    The noise level used for the realistic radio--like galaxies is $\sigma_{uv} = 0.01$ (which is not directly comparable to the parametric dataset as the parametric and autoencoded galaxies have different fluxes), resulting in an S/N range of $20$ to $60$.
    The S/N distribution is shown in Figure 3 for both the parametric and the realistic galaxies. The latter contain no galaxies of very low S/N (between $10$ and $20$), however, the distribution is consistent with that of the parametric galaxies.
    An example of a COSMOS galaxy adapted to a radio-like format is shown in \autoref{fig:cosmos_radio_like}, which displays the absolute values of the gridded visibilities and its Fourier counterpart, the galaxy dirty image. 
    
\begin{figure*}
                \centering
                \includegraphics[width=\linewidth, trim={.4cm .3cm .2cm 1cm},clip]{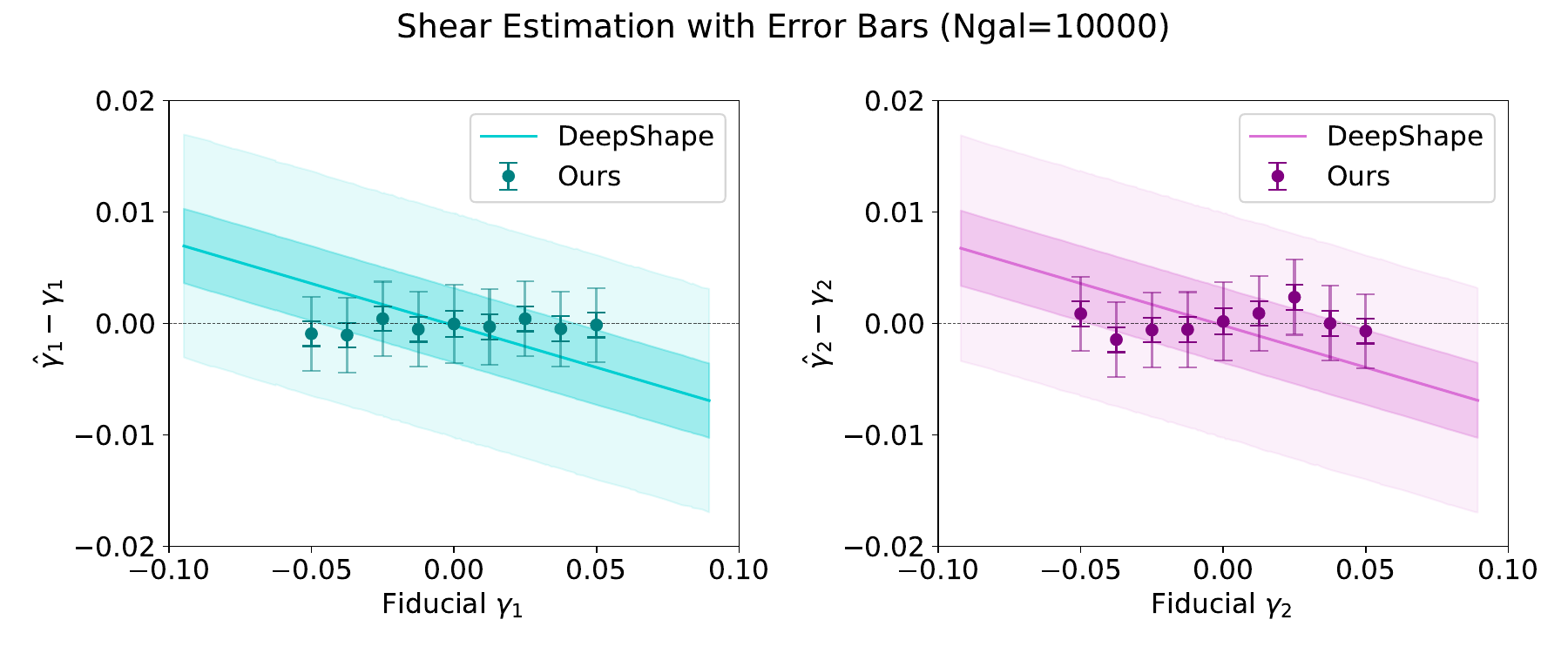}
                \caption{Parametric galaxies --- parametric forward model: shear estimation error as a function of the fiducial shear. The first component of the shear, $\gamma_1$, is shown on the left-hand side of the figure and the second component, $\gamma_2$, on the right-hand side. Our results are represented by dots, with error bars showing the $1\sigma$ and $3\sigma$ intervals. For comparison, a linear interpolation of the results from DeepShape’s Fig. 9 is shown, together with the corresponding $1\sigma$ and $3\sigma$ intervals. }
                \label{fig:results_vs_deepshape}
        \end{figure*}
        
\section{Results}
\label{sec:results}

    In this section, we present the results of the shear posterior estimation. First, we show the results obtained using the parametric model and compare them with other state-of-the-art methods for shear measurement. 
    We then demonstrate the model bias that can occur when considering galaxies with more complex morphologies than those represented by the adopted generative model. 
    Finally, we address this bias by using a more complex generative model, which is capable of representing the morphology of the observed galaxies.
        
    \subsection{Parametric galaxies}
    \label{sec:results_parametric}
        We estimate the shear posterior distribution using the forward model and the MCLMC sampling algorithm for different fiducial shear values. We selected nine ground-truth points on the $(\gamma_1,\gamma_2)$ plane to obtain equidistant values between $-0.05$ and $0.05$ for each shear component. For each fiducial shear, we simulated the visibilities of $10\,000$ parametric galaxies, as described in \autoref{sec:data_param}. 
        The number of galaxies used corresponds to the galaxy density expected for the first phase of SKA (SKA1), $n_{\rm gal} \sim 2.7\,{\rm arcmin}^{-2}$ \citep{Harrison_2016}, over an area of $1$ deg$^2$, approximately the primary beam of the MeerKAT antennas, which gives roughly $10\,000$ galaxies.
        In this test, we use the GalSim-based Spergel generative model, whose morphology is consistent with the simulated observations. The sampled parameters and their prior distributions are listed in \autoref{tab:parameter_prior_spergel}.
        
        Since the MCLMC sampling algorithm requires the gradients of the forward model to be calculated for each simulated galaxy, we must process the $10\,000$ galaxies in smaller batches due to GPU memory constraints. 
        We use a single Nvidia Tesla V100 GPU with 32 GB of memory, and the forward modelling of the galaxy images, together with the gradient computation graphs, can easily exceed the available memory for a few hundred galaxies.
        We therefore sample the posterior distribution using batches of 100 galaxies at a time (giving a total of 100 batches). 
        Sampling takes around $10$ minutes, using $10$ chains in parallel, taking $200\,000$ samples and yielding a total ESS of around $4\,000$ samples for the shear parameters. As we divide the dataset into $100$ batches of $100$ galaxies each, we can parallelise the runs across multiple GPUs, achieving a performance of approximately $60$ ms per galaxy.  
        In this paper, we do not seek to optimise the performance of the algorithm, but simply to demonstrate its ability to produce an unbiased estimate of cosmic shear.
        After obtaining the samples for a given batch, we estimate the posterior density by fitting a GMM. Finally, we combine the GMMs from every batch to estimate the posterior density for the entire dataset. 
        The posterior distributions have a single mode and a Gaussian-like appearance, so they are combined by analytically multiplying their densities. 
        Since the posteriors are likelihood-dominated,the precision of the prior accounting for approximately $1\%$ of the posterior precision, we directly multiply the posteriors without subtracting the prior.

        \begin{table}
            \centering
            \caption{Sampled parameters with their respective prior distributions and sizes. The first four parameters are sampled for each observed galaxy, whereas the shear parameter, $\gamma = (\gamma_{1}, \gamma_{2})$, is unique and shared by all observations within the batch. }
            \begin{tabular}{@{}cccc@{}}
            \toprule
            Parameter      & Prior distribution             & Size & Units \\ \midrule
            hlr            & LogisticNormal$(0.1,\, 3.0)$   & $(N_{\rm gal},)$ & arcsec\\
            $\nu$          & LogisticNormal$(-0.7,\, 1.0)$  & $(N_{\rm gal},)$ & - \\
            flux           & LogisticNormal$(0.03,\, 0.25)$ & $(N_{\rm gal},)$ & mJy \\
            $(\rm{e}_{1},\rm{e}_{2})$ & $\mathcal{N}(0, \, 0.2)$       & $(N_{\rm gal}, 2)$ & - \\
            $(\gamma_{1}, \gamma_{2})$ & $\mathcal{N}(0, \, 0.1)$       & $(1,2)$ & - \\ \bottomrule
            \end{tabular}
            \label{tab:parameter_prior_spergel}
        \end{table}

        \Autoref{fig:results_vs_deepshape} shows the results of the shear estimation. In each panel, we present the estimation error for each shear component as a function of the different fiducial values. For each shear value, we calculate the mean and standard deviation of the GMM derived from the $10\,000$ observations. The error bars represent the $1\sigma$ and $3\sigma$ intervals. 
        For reference, we show the shear estimation error of the DeepShape method \citep{tripathi2025deepshape}, which uses convolutional neural networks to estimate the individual ellipticity of each galaxy and then estimates the shear as the average of these ellipticities. 
        The results of the galaxy-by-galaxy ellipticity estimates for various fiducial shear values were provided by the authors of DeepShape. 
        We have reproduced the Figure 9 of their paper, simplifying the data points to a linear interpolation to improve readability. 
        In \citet{tripathi2025deepshape}, the authors also use $10\,000$ images of parametric galaxies with an equivalent radio PSF, size and flux distribution. 
        The difference between their dataset and ours lies in the type of parametric profile used, as DeepShape simulates galaxies using Sérsic profiles. 
        However, we consider that this difference has no significant impact on the results, as both models produce similar galaxy morphologies for the chosen parameter ranges, and use the same $uv$ coverage, and number of galaxies. 
        Comparing the two methods, we note that our shear estimate is more accurate, with an uncertainty approximately three times smaller for the same number of galaxies. Furthermore, our method shows no apparent signs of significant multiplicative bias.

    \subsection{Posterior calibration}
        Sampling the posterior probability naturally enables uncertainty quantification over the shear estimates.
        However, the process does not guarantee that the estimated density is well calibrated. 
        To assess the quality of the estimated shear posterior, we perform a probability coverage test for a single reference shear value. The selected shear is $\gamma_{\rm GT}=(-0.05, 0.05)$. 
        Since we are dealing with simulations, we know the expected value of the posterior probability; we therefore carry out an analysis based on the distribution of the high posterior density (HPD) credible regions \citep{cook2006validation, talts2018validating}.
        For the chosen ground-truth shear value, we estimate the posterior density using our framework a total of $N$ times.
        Then, for each estimated posterior $\hat{p}_n(\gamma|\mathcal{D})$, we find the interval of highest density that contains the ground-truth shear value,
        \begin{equation}
            I_n^{\rm HPD} = \{\gamma \in \mathbb{C} :\hat{p}_n(\gamma|\mathcal{D}) \ge \hat{p}_n(\gamma_{\rm GT}|\mathcal{D}) \}.
        \end{equation}
        We then compute the credible level $C_n$ by integrating the posterior density over this interval, as follows
        \begin{equation}
            C_n = \int_{\gamma \in I_n^{\rm HPD}} \hat{p}_n(\gamma|\mathcal{D}) d\gamma.
        \end{equation}
        The theoretical distribution of $C_n$ for a well-calibrated posterior is uniform between 0 and 1. Therefore, for a given nominal confidence level $\alpha \in [0,1]$, the fraction $K/N$ of posteriors for which the ground-truth shear falls within the $\alpha$-credible region, that is, $C_n \le \alpha$,  should be equal to $\alpha$. The number of posteriors for which the ground truth value falls inside the $\alpha$-credible region, $K$, is distributed as a ${\rm Binomial}(N,\alpha)$, with mean $N\alpha$ and variance $N\alpha(1-\alpha)$. Therefore, the expected value of the ratio $K/N$ is $\alpha$, and its standard deviation is
        \begin{equation}
            \sigma_{K/N} = \sqrt{\frac{\alpha (1-\alpha)}{N}}.
        \end{equation}

\begin{figure}
                \centering
                \includegraphics[width=\linewidth, trim={.1cm .5cm .1cm .2cm},clip]{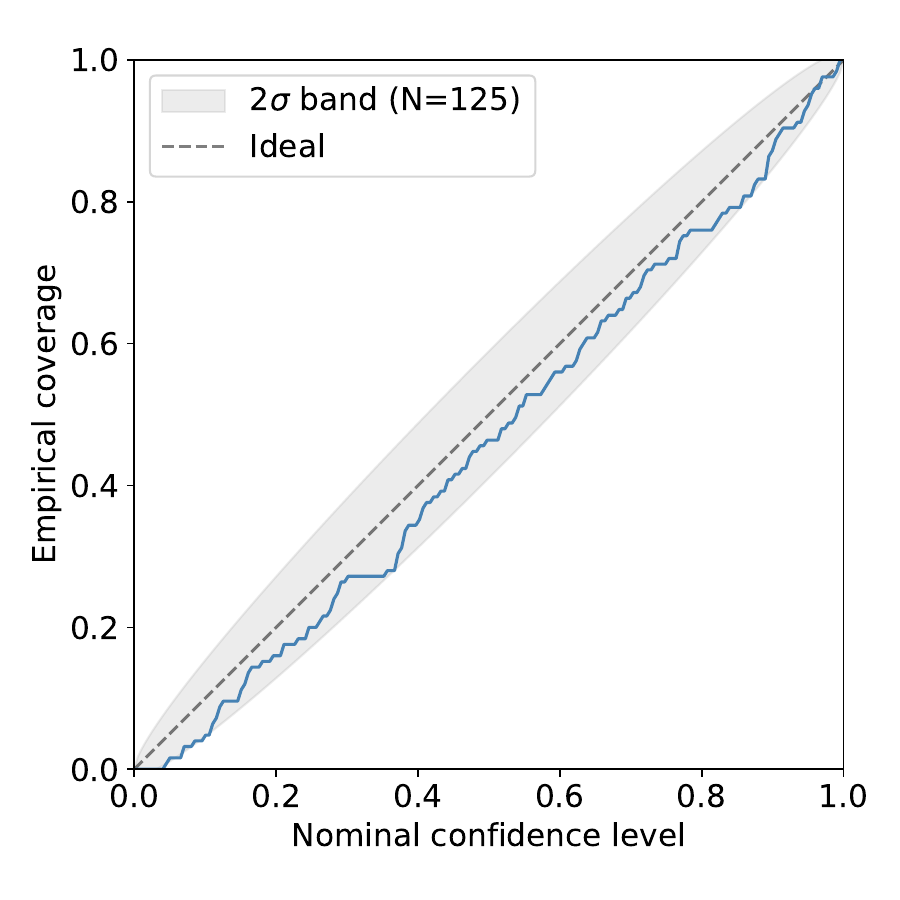}
                \caption{Parametric galaxies --- parametric forward model: ratio of coverage probability as a function of the confidence level $\alpha$. The grey band shows the binomial standard deviation from the ideal coverage on the $\alpha :\alpha$ diagonal.  }
                \label{fig:coverage}
        \end{figure}

        We performed the coverage test for $N=125$ posterior estimates and present the results in \autoref{fig:coverage}, which shows the empirical coverage ratio for different values of the credibility level $\alpha$. The diagonal line $\alpha:\alpha$ represents the ideal coverage, and we have outlined the $2\sigma$ band in grey. We note that the empirical coverage, shown in blue, is consistent with the theoretical coverage, indicating that the estimated posteriors are properly calibrated.
        However, it should also be noted that it systematically lies below the ideal coverage, which we will analyse in the future with a larger number of simulations.
        
    \subsection{Realistic galaxies}
        In this section, we test the forward model on galaxies with more complex morphologies, evaluating shear estimation in a scenario where the generative model is unable to capture the morphological complexity of the observations. 
        We use the COSMOS real galaxy dataset adapted to radio interferometric observations, which is described in \autoref{sec:data_cosmos}.  
        We estimate the shear posterior for a total of $10\,000$ whitened COSMOS galaxies. As we are solely seeking to demonstrate the existence of model bias in the case of morphological mismatch, we perform this test for a single fiducial ground-truth shear value of $\gamma_{\rm GT}=(-0.05, 0.05)$.

        \subsubsection{Spergel generative model}
        First, we evaluate the forward model using the parametric generative model, which models galaxies using a Spergel profile. 
        We subdivide the dataset into batches of $100$ galaxies to ensure they fit in the GPU memory, and we sample the posterior using MCLMC, as described in \autoref{sec:sampling}.
        Given that the size and range of the COSMOS profiles do not follow the same distribution as the T-RECS dataset, we use different priors, which are shown in \autoref{tab:parameter_prior_spergel_cosmos}.
        Once the posterior samples have been obtained, we estimate the density using a GMM. 
        Finally, we analytically combine the posteriors across the multiple runs.
        \begin{table}
            \centering
            \caption{Parameters and priors of the Spergel model used with the whitened COSMOS real galaxies dataset. The first four parameters are sampled for each observed galaxy, whereas the shear parameter, $\gamma = (\gamma_{1}, \gamma_{2})$, is unique and shared by all observations within the batch. }
            \begin{tabular}{@{}cccc@{}}
            \toprule
            Parameter      & Prior distribution             & Size & Units \\ \midrule
            hlr            & LogisticNormal$(0.1,\, 5.0)$   & $(N_{\rm gal},)$ & arcsec\\
            $\nu$          & LogisticNormal$(-0.7,\, 1.0)$  & $(N_{\rm gal},)$ & - \\
            flux           & LogisticNormal$(0.005,\, 1.0)$ & $(N_{\rm gal},)$ & mJy \\
            $(\rm{e}_{1},\rm{e}_{2})$ & $\mathcal{N}(0, \, 0.2)$       & $(N_{\rm gal}, 2)$ & - \\
            $(\gamma_{1}, \gamma_{2})$ & $\mathcal{N}(0, \, 0.1)$       & $(1,2)$ & - \\ \bottomrule
            \end{tabular}
            \label{tab:parameter_prior_spergel_cosmos}
        \end{table}
        The posterior shear estimate from $10\,000$ galaxies using the Spergel generative model is shown in violet in \autoref{fig:shear_cosmos}.
        The black cross indicates the ground truth shear value. The constraining power of the parametric model is similar to the results shown in \autoref{sec:results_parametric}, however, in this case we observe a considerable bias, larger than $10\sigma$.
        This bias is to be expected, as Spergel profiles are unable to capture the complex galaxy morphologies present in the COSMOS dataset. 

        \subsubsection{NN-based generative model}

        To address this bias, we consider the AE-based generative model, which does capture complex galaxy morphologies. We repeated the shear inference using $10\,000$ radio-adapted COSMOS galaxies. The galaxy parameters in this model correspond to the latent vector $z \in \mathbb{R}^{4\times4}$ of the AE. In order to sample in the AE latent space, we use the normalising flow as described in \autoref{sec:generative_model}. The sampled parameters and their priors are listed in \autoref{tab:parameter_prior_ae_cosmos}.
        \begin{table}
            \centering
            \caption{Parameters and priors of the NN-based generative model used with the whitened COSMOS real galaxies dataset. The shear parameter, $\gamma = (\gamma_{1}, \gamma_{2})$, is unique and shared by all observations within the batch. The latent vectors are obtained through the normalising flow, $z = f(u)$.}
            \begin{tabular}{@{}cccc@{}}
            \toprule
            Parameter      & Prior distribution             & Size & Units \\ \midrule
            $u$            & $\mathcal{N}(0,\, I)$   & $(N_{\rm gal},4,4)$ & -\\
            $(\gamma_{1}, \gamma_{2})$ & $\mathcal{N}(0, \, 0.1)$       & $(1,2)$ & - \\ \bottomrule
            \end{tabular}
            \label{tab:parameter_prior_ae_cosmos}
        \end{table}
        Sampling with MCLMC requires calculating the gradients of the forward model with respect to the sampled parameters. 
        Given that the forward model comprises a NN-based flow and generative model, such gradients require considerable GPU memory and are slower to compute than in the case of a simple parametric model.
        In order to run the MCLMC sampling of $100$ galaxies on a single Nvidia Tesla V100 GPU with this model, we carried out various performance optimisation efforts.

        \begin{figure}
                \centering
                \includegraphics[width=\linewidth, trim={.1cm .0cm -.5cm -.5cm},clip]{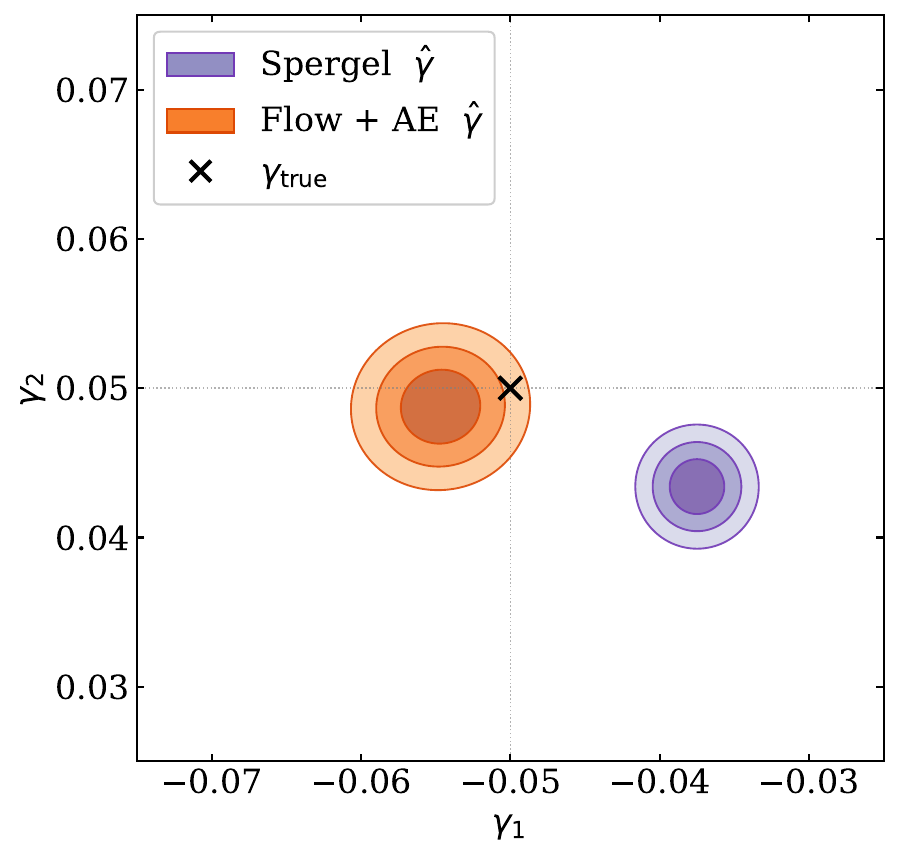}
                \caption{Realistic galaxies --- parametric and autoencoder forward models: shear posterior estimate for the radio adapted COSMOS real galaxies. In violet, the shear estimate using the parametric galaxy model for a total of $10\,000$ galaxies is shown. The ground-truth shear is indicated by the black cross. As expected, a considerable bias is observed in the shear estimate due to the model morphology mismatch. In orange, the shear estimate using the AE-based generative model for a total of $10\,000$ galaxies is shown.}
                \label{fig:shear_cosmos}
        \end{figure}

        The simplest solution is to use a smaller generative model. However, it is difficult to successfully train a small autoencoder model with few parameters that is able to capture the complex structure of the target images. Then, the chosen approach is to use a technique known as knowledge distillation \citep{bucilua2006model, hinton2015distilling}. This technique involves using a pre-trained model (the decoder part of the AE) as a teacher for a considerably smaller model, the student. This task is simpler than training the entire AE model, as the student model only needs to learn the mapping from the latent space to the surface brightness profile of the galaxies and not the full encoding and decoding task.
        We used a student model with the same number of layers but a depth (number of channels) that is a quarter of that of the teacher model, reducing the number of parameters from $18$ million to $790$ thousand. This significantly reduces the memory footprint of the gradients and also speeds up sampling, as the forward pass of the distilled model is 10 times faster.
        To speed up the forward pass even further, we recompiled the distilled model using half-precision, with \texttt{float16} instead of \texttt{float32}. 
        In the tests we carried out, the reduction in numerical precision did not significantly impact simulated visibilities ($<0.02\%$ relative error) or the computed gradients ($<1\%$ relative error).
        Finally, we limit the padding size used by JAX-GalSim to interpolate galaxy images and plot them in Fourier space.
        These optimisations allow us to use MCLMC sampling with batches of $100$ galaxies, limiting the number of parallel chains to four. We took a total of $5\,000$ samples, resulting in an ESS of around $200$ samples for the shear parameters.
        Sampling takes $50$ minutes, approximately $300$ ms per galaxy, by running the $100$ runs in parallel across multiple GPUs. 

        Once we have sampled the posterior for each batch of 100 galaxies, the posteriors from each run are combined by analytically multiplying the estimated densities.
        The results of the shear posterior inference are shown in orange in \autoref{fig:shear_cosmos}. 
        As can be seen in this case, the shear estimate is not biased, with the ground-truth shear value lying close the edge of the 2$\sigma$ region.
        The ability of the NN-based generative model to represent the observations makes it possible to address the model bias observed in the parametric case. 
        We also note that the posterior estimate is slightly larger than in the parametric case. This is because the NN-based generative model has greater flexibility, and there are more combinations of latent parameters and shear that adequately represent the observed data.

    \subsection{Scalability and performance}
    Surveys such as SKA2 promise to observe a high galaxy density, over $10$ ${\rm arcmin}^{-2}$ \citep{Harrison_2016}. Given their extensive sky coverage, it is expected that several trillion galaxies will be detected.
    Therefore, producing shear maps of large regions of the sky requires processing a very large number of galaxies.
    Estimating the full shear posterior using probabilistic sampling techniques allows us to quantify the uncertainty of the estimate and marginalise over the latent parameters; however, it is slow compared to point estimate methods. 
    Moreover, estimating the full posterior becomes considerably slower the more complex the forward model is. As an example, the proposed parametric model takes around $10$ minutes to get $2\,500$ effective shear samples for a batch of $100$ galaxies, whilst the NN-based model takes $50$ minutes for a much smaller number of effective samples.
    This performance limitation is due not only to the time taken by the model to simulate the observations, but also to the time taken to compute the gradients of the model with respect to the sampled parameters. 
    With the aim of assessing the scalability of our method for future surveys, we investigate an alternative to full posterior estimation. 
    In this section, we carry out a test in which we use the forward model to predict a point estimate of the shear using the same observations as those considered in the previous case. 
    The approach chosen as the point estimation method consists of posterior density maximization (MAP).
    
    Unlike the MAP optimisation used to initialise the sampling algorithm, in this case we want to ensure convergence to a maximum of the posterior density, where in the previous case where it was sufficient to be close to a likely region in order to carry out the sampling. 
    We have observed experimentally that initialising the latent parameters to zero yields better results. We believe this is due to the flexibility of the AE model in representing the data; consequently, initialising at a random point in the latent space can lead to convergence at local maxima.
    We initialise the shear randomly, sampling it from the prior. We then perform four MAP optimisations in parallel to verify that they converge to the same shear value. 
    We use a learning rate $lr =0.05$ to ensure convergence, and applied a $2\times$ factor to the $lr$ related to the shear parameters. The aim is for the optimisation to adjust the shear parameters rather than searching for latent parameters that explain the observations for a given initial shear. 
    We run the MAP optimisation for a total of $1\,500$ gradient descent steps using the Adam optimiser from the \texttt{optax} library.
    As in the previous tests, due to GPU memory limitations, the dataset is divided into batches of $100$ galaxies, obtaining $100$ shear point estimates. The $100$ runs are executed in parallel, taking a total of $6$ minutes, compared to almost $1$ hour for the full posterior. 
    The MAP estimation performance can be further optimised by running just one MAP per run, reducing the estimation time by a factor of four, resulting in a processing time of $9$ ms per galaxy.
    
\begin{figure}
        \centering
        \includegraphics[width=\linewidth, trim={.1cm .0cm -.5cm -.5cm},clip]{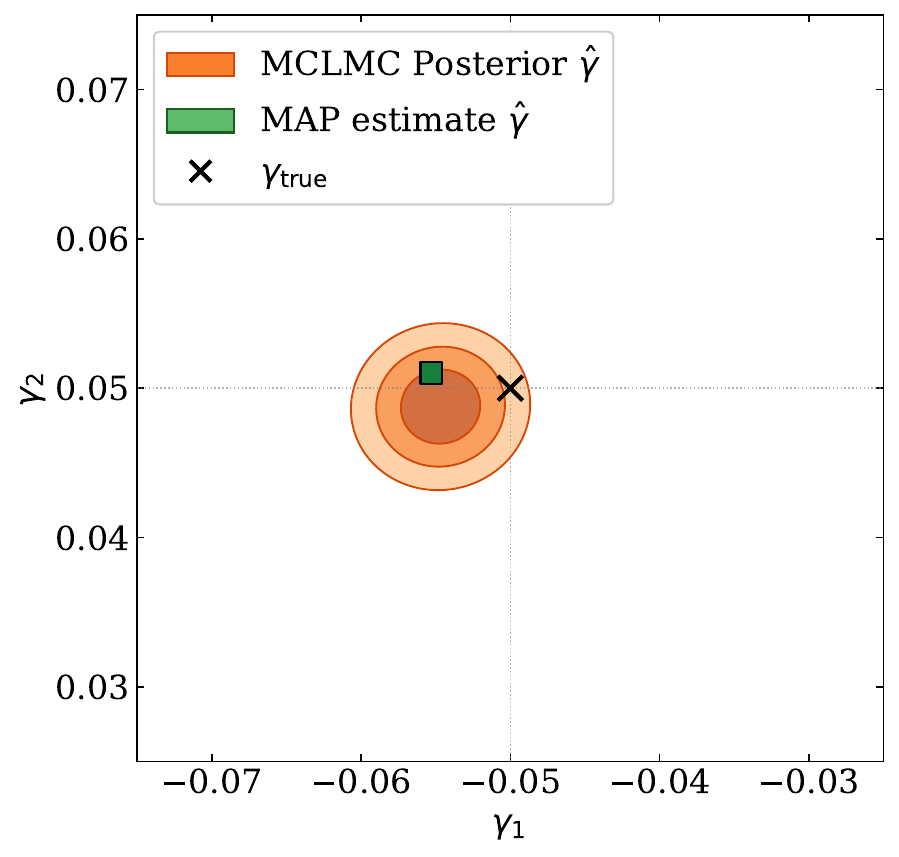}
        \caption{Realistic galaxies --- autoencoder forward model: MAP point estimate of the cosmic shear for $10\,000$ radio adapted COSMOS real galaxies, using the NN-based forward model.}
        \label{fig:results_map}
    \end{figure}

    The results obtained from the $100$ runs are combined, and the mean is shown in \autoref{fig:results_map} in green, alongside the full posterior estimate from the previous test. 
    We note that the MAP estimate agrees with the estimated posterior and is as close to the ground truth as the maximum of the estimated posterior. 
    The MAP estimate does not coincide with the maximum of the estimated posterior distribution, as the posterior shown here is marginalised with respect to the galaxy parameters, whereas the MAP is optimised over the joint posterior distribution.
    One of the main limitations of this method is that we do not have a direct quantification of the uncertainty in the estimated shear. 
    Furthermore, as we are not marginalising over the forward model's galaxy parameters, any degeneracy between this parameters and the shear will potentially introduce a bias.
    Finally, in \cite{refregier2012noise}, the authors show that maximum likelihood estimation (MLE) methods for template fitting galaxy shapes from optical images suffer from noise bias, leading to biases in cosmic shear estimation.
    Our scenario is different, as the likelihood is modelled in Fourier space and therefore the noise does not interact in the same way with the shape of the galaxy; however, further tests must be carried out to guarantee an unbiased point estimate of the shear.

\section{Conclusions and perspectives}
\label{sec:conclusions}

    \subsection{Conclusions}
    In this paper we propose a shape-measurement-free method for estimating cosmic shear, based on a forward model for simulating galaxy observations.
    The model consists of three components: a generative model of galaxy images, the distortion introduced by cosmic shear, and the instrumental response of the radio interferometer. For a given combination of galaxy parameters, cosmic shear, and radio PSF, the model simulates the radio visibilities in the $uv$ plane, which are then compared with the observations. 
    Using a probabilistic sampling algorithm, the posterior density of the shear parameters can be estimated conditioned on a set of observed galaxies. Since galaxy shape and cosmic shear are inherently degenerate, observations of multiple galaxies from the same sky region are used jointly, under the assumption that they all share the same shear value.
    We employ two types of generative models: one based on the parametric Spergel surface brightness profile, whose morphological parameters are the half-light radius (hlr), flux, Spergel index $\nu$, and the intrinsic ellipticity of the galaxy; and a NN-based generative model, consisting of the decoder part of an autoencoder architecture trained on real galaxy images (COSMOS) from the Hubble Space Telescope, which encodes galaxy morphology directly in the latent space of the autoencoder.
    Our forward model is implemented in JAX and is fully differentiable, which allows the use of gradient-based sampling methods such as MCLMC, which are highly efficient when sampling a large number of parameters.

    Since real radio observations with sufficient resolution and galaxy density for WL studies are not yet available, we test our method on simulated radio galaxy visibilities. In a first evaluation, we use parametric profiles in real space with a parameter distribution based on the T-RECS catalogue of radio star-forming galaxies, as employed in other state-of-the-art works on galaxy shape measurement in radio interferometry. 
    We use a radio PSF based on the SKA-Mid antenna layout for an $8$-hour observation at zenith.
    We demonstrate that our framework is capable of estimating the shear for a total of $10\,000$ simulated galaxy visibilities, and compare the results with other state-of-the-art methods such as DeepShape. Our forward model provides an estimate with lower uncertainty and no appreciable bias. Furthermore, we assess the quality of the estimated posterior by performing a coverage test, which shows that the posterior is correctly calibrated.

    We then test the parametric forward model on simulations of galaxies with complex morphologies. For these simulations, we use an autoencoder trained on real galaxy images from the HST COSMOS survey to generate realistic surface brightness profiles.
    We simulate the visibilities of $10\,000$ galaxies sheared by the same amount, and estimate the shear using the parametric forward model. We show that the resulting shear estimate is biased, due to the inability of the parametric model to capture complex galaxy morphologies. To address this, we repeat the same test replacing the parametric forward model with the NN-based generative model, and show that accurately representing the morphology of the target galaxies leads to an unbiased cosmic shear estimate.

    Finally, we used the NN-based forward model to perform a point estimate of the shear using a MAP optimisation. Obtaining a point estimate is considerably faster than a full posterior estimate but lacks uncertainty quantification. 
    Furthermore, it may potentially be biased due to degeneracies between the galaxy parameters and the shear, in which case an additional calibration step will be required.
    We show that we were able to use the proposed forward model to obtain a point estimate of the shear that is consistent with the sampled posterior distribution. This method is up to $30$ times faster than sampling the full posterior. Further tests are needed to properly assess the bias of the point estimate.

    \subsection{Future work}

    In this work we demonstrate the ability to obtain a posterior estimate of cosmic shear from radio data by forward modelling the observed galaxies. To this end, we consider several simplifications. We do not address galaxy detection, which may introduce a bias due to the preferential selection of galaxies with certain characteristics. Galaxies are simulated individually, centred on postage stamps, rather than simulating the $uv$ samples for all the galaxies distributed across a single FOV. We use the same radio PSF for all observations. We do not account for the spectral energy distribution (SED) of the observed galaxies when simulating visibilities across different frequency channels.
    Furthermore, we do not take into account the possible intrinsic alignment of background galaxies with their hosting LSS, which may correlate the shapes of galaxies even in the absence of shear. These considerations remain to be explored in future work.   
    This work is a demonstration of the method using radio simulations, and represents an initial step towards a shear inference environment (SHINE\footnote{\url{https://github.com/CosmoStat/SHINE}}) that we are currently developing. Ultimately, SHINE will forward model galaxy observations across the entire FOV, jointly fitting the total number of galaxies and their positions, using a spatially and spectrally varying PSF, and accounting for the spectra of the observed galaxies, while simultaneously estimating the shear and the morphological and physical parameters of the galaxies, such as their metallicity or star formation rate, which are related to their spectra. SHINE will make use of both radio and optical observations, with the ability to jointly estimate the posterior of the target parameters using data from multiple surveys.

\begin{acknowledgements}
      This work was granted access to the HPC resources of IDRIS under the allocation AD010414104R3 and AD011012983R4 made by GENCI.
      This work was supported by the European Community through the grant ARGOS (contract no. 101094354).
      This work was supported by the TITAN ERA Chair project (contract no. 101086741) within the Horizon Europe Framework Program of the European Commission, and the Agence Nationale de la Recherche (ANR-22-CE31-0014-01 TOSCA).
      We would like to thank Priyamvad Tripathi for sharing the data used to reproduce the DeepShape results plot.
\end{acknowledgements}

\begin{dataavailability}
The scripts and notebooks for reproducing the results and figures in this paper are available at: \\
\tiny\href{https://github.com/CentofantiEze/Radio_WL_Generative_Forward_Model}{https://github.com/CentofantiEze/Radio\_WL\_Generative\_Forward\_Model}. 
The trained models and metrics are available at \href{https://zenodo.org/records/21131669}{https://zenodo.org/records/21131669}.
\end{dataavailability}

%
\bibliographystyle{aa} 
\bibliography{references} 

\clearpage




\end{document}